\documentclass[pre,aps,onecolumn,groupedaddress]{revtex4-1}
\usepackage{amssymb,epsfig,amsmath,bm,subfigure,graphicx,textcomp,url,float,color,cases}
\usepackage[normalem]{ulem}

\begin{document}
\title{Large deviations of surface height in the $1+1$-dimensional Kardar-Parisi-Zhang equation: exact long-time results for $\lambda H<0$}

\author{Pavel Sasorov} \affiliation{Keldysh Institute of Applied Mathematics, Moscow, 125047, Russia}
\author{Baruch Meerson} \affiliation{Racah Institute of Physics, Hebrew University of Jerusalem, Jerusalem 91904, Israel}
\author{Sylvain Prolhac} \affiliation{Laboratoire de Physique Th\'{e}orique;  Universit\'{e} de
Toulouse, UPS, CNRS, Toulouse, France}

\begin{abstract}
We study atypically large fluctuations of  height $H$ in the 1+1-dimensional Kardar-Parisi-Zhang (KPZ) equation at long times $t$,
when starting from a ``droplet" initial condition.
We derive exact large deviation function of height for $\lambda H<0$, where $\lambda$ is
the nonlinearity coefficient of the KPZ equation. This large deviation function describes a crossover
from the Tracy-Widom distribution tail  at small $|H|/t$, which scales as $|H|^3/t$,
to a different tail at large $|H|/t$, which scales as $|H|^{5/2}/t^{1/2}$. The latter tail exists at all times $t>0$.
It was previously obtained
in the framework of the optimal fluctuation method. It was also obtained at short times from exact representation of the complete height statistics. The crossover between the two tails, at long times, occurs at $|H|\sim t$
as previously conjectured. Our analytical findings are supported by numerical evaluations using exact representation of the complete height statistics.

\end{abstract}

\maketitle
\tableofcontents

\section{Introduction}

The celebrated Kardar-Parisi-Zhang (KPZ) equation \cite{KPZ}  defines an important universality class
of non-equilibrium surface growth \cite{HHZ,Krug,Corwin,QS,S2016}.
In $1+1$ dimension this equation,
\begin{equation}\label{KPZoriginal}
\partial_{t}h=\nu \partial^2_{x}h+(\lambda/2)\left(\partial_{x}h\right)^2+\sqrt{D}\,\xi(x,t),
\end{equation}
describes the evolution of the interface height  $h(x,t)$ driven by a Gaussian white noise  $\xi(x,t)$ with zero mean
and covariance
\begin{equation}\label{LD040}
\left\langle\xi(x,t)\xi(x^\prime,t^\prime)\right\rangle=\delta\left(x-x^\prime\right)\, \delta\left(t-t^\prime\right)\,.
\end{equation}
The diffusion term describes relaxation of the interface, whereas the nonlinear term breaks the symmetry $h \leftrightarrow -h$ in an essential way.
At long times the interface width, governed by Eq.~(\ref{KPZoriginal}), grows as $t^{1/3}$, whereas the horizontal correlation length grows as $t^{2/3}$. These power laws -- the hallmarks of the KPZ universality class -- were confirmed in experiments \cite{experiment1}. In the recent years the focus of interest in the KPZ equation  shifted toward a more detailed characterization of the fluctuating interface, such as the complete one-point probability distribution  ${\mathcal P}_t(H)$ of height $H$ at a specified time $t$ at a specified point in space \cite{Corwin,QS,S2016}. For the KPZ equation in $1+1$ dimension several groups derived exact representations for a generating function of  ${\mathcal P}_t(H)$
at any $t>0$. These remarkable results have been obtained for three classes of initial conditions (and for some
combinations of them): flat interface \cite{CLD}, ``droplet" \cite{SS,CDR,Dotsenko,ACQ,Corwin}, and Brownian, stationary interface \cite{IS,Borodinetal}.  In the long-time limit, and for typical fluctuations, ${\mathcal P}_t(H)$ converges to the Tracy-Widom (TW) distribution for the Gaussian orthogonal ensemble (GOE)  \cite{TWGOE} for the flat interface, to the TW distribution for the Gaussian unitary ensemble (GUE) \cite{TW} for the droplet, and to the Baik-Rains distribution \cite{BR} for the stationary interface. A series of ingenious experiments with liquid-crystal turbulent fronts fully confirmed these long-time results for typical fluctuations \cite{experiment2}.

Less is known about \emph{large deviations}, that is atypically large fluctuations of the surface height, which are described by the far tails of ${\mathcal P}_t(H)$.  Extracting
these tails from the exact representations requires considerable effort. As of present, there have been only two attempts in this direction, made by Le Doussal \textit{et al.} for the droplet initial condition: for long \cite{DMS} and short \cite{DMRS} times. We will comment on their results as we proceed.

Given the difficulties in extracting the tails from the exact representations, one can look for alternatives that would directly probe the far tails of ${\mathcal P}_t(H)$. One such alternative has existed long before the exact representations for the height distribution of the 1+1 dimensional KPZ equation were found.
It appears in different areas of physics under different names: the optimal fluctuation method (OFM), the instanton method, the weak noise theory, the
macroscopic fluctuation theory, etc. In the context of the KPZ equation the OFM was employed in Refs.  \cite{Fogedby1998,Fogedby1999,Fogedby2009,KK2007,KK2008,KK2009,MKV,KMS,JKM}. The crux of the method is
a saddle-point evaluation of the path integral for the KPZ equation conditioned on a specified large deviation.
Correspondingly, it requires a small parameter (hence the term ``the weak noise theory"). In $1+1$ dimension this small parameter turns out to be proportional to $t^{1/2}$ \cite{KK2007,KK2008,KK2009,MKV,KMS,JKM}. As a result, at short times,
the OFM correctly describes the \emph{complete} large-deviation function (LDF) of the interface height. For a whole class of initial conditions, including the three initial conditions described above, the tails of this short-time LDF,  determined with the OFM,
scale as $|H|^{3/2}/t^{1/2}$ (for $\lambda H>0$) and $|H|^{5/2}/t^{1/2}$ (for $\lambda H<0$). For the droplet initial
condition these tails agree with the corresponding short-time tails obtained by Le Doussal \textit{et al.} \cite{DMRS}.

When is the OFM applicable at long times? A necessary condition is that the LDF of height,
predicted by the OFM (it is equal to the action of the classical field theory emerging in the OFM) is much larger than unity \cite{KK2007,KK2008,MKV,KMS,JKM}.   At arbitrarily long but finite times this condition is always satisfied sufficiently far in the tails of ${\mathcal P}_t(H)$. It is possible, however, that a dominant contribution to ${\mathcal P}_t(H)$ comes from \emph{non-saddle-point} histories $h(x,t)$. This is indeed what happens at long times in the $\lambda H<0$ part of ${\mathcal P}_t(H)$ for the KPZ equation. At small $|H|/t$ the
GOE TW tail, the GUE TW tail and the Baik-Rains tail all scale as $|H|^3/t$, and this is much smaller than $|H|^{5/2}/t^{1/2}$
predicted by the OFM.  The situation is reversed at large  $|H|/t$. Therefore, it was conjectured in Refs. \cite{MKV,KMS,JKM} that, at $|H|\sim t$, each of the $|H|^3/t$ tails of the GOE TW, GUE TW and the Baik-Rains distributions crosses over to the corresponding $|H|^{5/2}/t^{1/2}$ tail that predicts a higher probability at large  $|H|/t$.

In this work we employ the exact representations  for the droplet initial condition \cite{SS,CDR,Dotsenko,ACQ,Corwin}
to derive exact $\lambda H<0$ LDF of height of the $1+1$-dimensional KPZ equation at long times. As we show, this LDF describes a smooth crossover between
the $|H|^3/t$ tail and  the  $|H|^{5/2}/t^{1/2}$ tail, in support of the above conjecture.

Here is how  the remainder of this paper is structured. In Sec. 2 we present the governing equations and the mathematical formulation of the problem. The problem is solved in Sec. 3. In Sec. 4 we discuss the properties of the LDF of height at $\lambda H<0$. Section 5 presents  results of a numerical evaluation of the LDF.  Section 6 includes a brief summary and discussion.

\section{Governing equations}

Let us assume that $\lambda>0$, so that the $\lambda H<0$ is the left tail of  ${\mathcal P}_t(H)$ \cite{signlambda}.
Following Ref.~\cite{DMS}, we will use in this paper the units of distance $x_0=(2\nu)^3/(D\lambda^2)$, time $t_0=2(2\nu)^5/(D^2\lambda^4)$ and height $h_0=2\nu/\lambda$. In these units Eq.~(\ref{KPZoriginal}) has
$\nu=1$ and $\lambda=D=2$ with the noise covariance~(\ref{LD040}). We consider the ``droplet" initial condition, conveniently represented by the $L\to 0$ limit of parabolic interface \cite{KMS}:
\begin{equation}\label{LD060}
h(x,0)= - \frac{x^2}{L} .
\end{equation}
We will study the probability distribution  ${\mathcal P}_t(H)$  of the shifted height $H$ at the origin at time $t$,
\begin{equation}\label{defH}
H:=h(x=0,t)+\frac{t}{12}+\frac{b t}{\delta} .
\end{equation}
The $t/12$ term is universal, whereas the $b t/\delta$ term is not: the coefficient $b=\mathcal{O}(1)$ depends
on the exact way of introducing a finite spatial correlation length $\delta$
(an ultraviolet cutoff) of the Gaussian noise \cite{Hairer}.

The exact representation for ${\mathcal P}_t(H)$ is the following~\cite{SS,CDR,Dotsenko,ACQ}.
Introduce
the generating function
\begin{equation}\label{LD050}
Q_t(s)=\left\langle \exp\left(-e^{H-t^{1/3}s}\right)\right\rangle\,,
\end{equation}
where the averaging is over the distribution ${\mathcal P}_t(H)$. This generating function is given
by a Fredholm determinant:
\begin{equation}\label{LD410}
Q_t(s)=\det \left[I-\hat{P}_s \hat{K}_t\hat{P}_s\right]\, ,
\end{equation}
where the kernel, corresponding to the operator $\hat{K}_t$, is
\begin{equation}\label{LD420}
K_t(x,x^\prime)=\int_{-\infty}^\infty \frac{\mbox{Ai}(x+v)\, \mbox{Ai}(x^\prime+v)}{1+e^{-t^{1/3}v}}\, dv\, ,
\end{equation}
$\hat{P}_s$ is the projector on the interval $[s,+\infty)$, and $\text{Ai}(\dots)$ is the Airy function.
Using this representation for typical fluctuations, $H=\mathcal{O}(t^{1/3})$, one obtains at long times ${\mathcal P}_t(H) = t^{-1/3} f(H/t^{1/3})$, where $f(s)$ is given by the GUE TW distribution \cite{SS,CDR,Dotsenko,ACQ,Corwin}. For the far right tail of ${\mathcal P}_t(H)$ one obtains \cite{DMS}:
\begin{equation}\label{farright}
- \ln {\mathcal P}_t(H) \simeq t \,\Phi_{+} \left(\frac{H}{t}\right), \quad \mbox{where}\quad\Phi_{+}(z)=\frac{4}{3} z^{3/2}.
\end{equation}
This leading-order asymptote coincides with the positive tail of the GUE TW distribution. It was derived from Eqs.~(\ref{LD050})-(\ref{LD420})  in Ref. \cite{DMS}. It was also obtained in Ref. \cite{KMS} by applying the OFM  to the KPZ equation
with the parabolic initial condition (\ref{LD060}) for arbitrary $L$, including the limits of $L\to 0$ and $L\to \infty$ \cite{different}.

The left tail of the GUE TW probability density, conjectured in \cite{TW} and proved in \cite{DIK2008}, is equal to
\begin{equation}\label{TW left tail}
f(s)\simeq2^{-47/24}\,e^{\zeta'(-1)}\,|s|^{15/8}\,e^{-\frac{|s|^{3}}{12}} ,
\end{equation}
where $\zeta(\dots)$ is the Riemann zeta function, and $\zeta'(-1)=-0.16542\dots$.
As we will see, the far left tail of ${\mathcal P}_t$ is quite different from the TW left tail (\ref{TW left tail}).
To determine the far left tail of ${\mathcal P}_t$, $|H|\sim t$, we will use an alternative exact representation, established
in Ref. \cite{ACQ}. The logarithm of the generating function $Q_t(s)$ can be expressed as
\begin{equation}\label{Qdirect}
  \ln  Q_t(s) = \int_s^{\infty}dr (s-r) \Psi_t(r),
\end{equation}
where
\begin{equation}\label{LD080}
\Psi_t(r)=\frac{t^{1/3}}{4}\int\limits_{-\infty}^{\infty} dv\, \text{sech}^2\,\left(\frac{t^{1/3} v}{2} \right)\, \left[q_t(r,v)\right]^2\,.
\end{equation}
The function $q_t(r,v)$ of three arguments $r,v$ and $t$ satisfies a nonlinear integro-differential equation,
\begin{equation}\label{LD090}
\partial_r^2 q_t(r,v)=\left[v+r+2\Psi_t(r)\right]\, q_t(r,v),
\end{equation}
subject to the boundary condition
\begin{equation}\label{LD096}
q_t(r,v)\bigr|_{r\to+\infty}\to \mbox{Ai}(r+v)\,.
\end{equation}
As this boundary condition is specified at plus infinity, we will need to know the behavior of $\Psi_t(r)$ in its \emph{right} tail, $r>0$. This behavior, at $s\sim t^{2/3} \gg 1$, has been recently established in Ref.~\cite{DMS}. Omitting
pre-exponential factors,
\begin{numcases}
{\!\!\Psi_t(s) \sim} e^{-\frac{4 s^{3/2}}{3}}, & $0<s\leq \frac{1}{4} \,t^{2/3}$,\nonumber \\
e^{-t \left(\frac{s}{t^{2/3}}-\frac{1}{12}\right)}, &$s \geq \frac{1}{4} \,t^{2/3}$. \label{Psilarger}
\end{numcases}
Our calculation of the LDF of height for the left tail,  $H<0$, relies
on an asymptotically exact solution of the problem~(\protect\ref{LD090}) and (\ref{LD096}), and asymptotic evaluation of the integrals~(\ref{Qdirect}) and (\ref{LD080}), at $t\gg 1$.

\section{Solution}
\label{WKB}

We are interested in the regime of $t\gg 1$ and $(-s)\gg 1$ (and, therefore, $-r\gg 1$).
Let us introduce the new variables
\begin{equation}\label{newvar}
X=\frac{r}{t^{2/3}}\quad\text{and}\quad V=\frac{v}{t^{2/3}}
\end{equation}
and make the  ansatz
\begin{equation}\label{ansatzPsi}
\Psi_t(r) =t^{2/3} g_t(X) \quad \mbox{and} \quad q_t(r,v) = t^{-1/6} \tilde{q}_t(X,V),
\end{equation}
where $g_t(X)>0$.
As it turns out, the function $g_t(X)$ is independent of $t$. We will not use this property in our calculations until later, but will suppress the subscript $t$ in  the function $g_t$ and in the related functions $U(X)$, $a(V)$ and $p(X,V)$ which we will introduce shortly. In the new variables Eq.~(\ref{LD090}) takes the form
\begin{equation}\label{LD090A}
\partial_X^2 \tilde{q}_t(X,V)+t^2 \left[-V-U(X)\right] \tilde{q}_t(X,V)=0\,,
\end{equation}
where $U(X)=X+2 g(X)$. The boundary condition (\ref{LD096}) becomes
\begin{equation}\label{E020a}
\tilde{q}_t(X,V)\bigr|_{X\to+\infty}\to t^{1/6} \mbox{Ai}\,[t(X+V)]\simeq \frac{\exp\left[-\frac{2}{3}t (X+V)^{3/2}\right]}{2\sqrt{\pi} \,(X+V)^{1/4}} ,
\end{equation}
where we have used the asymptotic of the Airy function for a large positive argument \cite{DLMF}.
In its turn, Eq.~(\ref{LD080}) can be rewritten as
\begin{equation}\label{LD080a}
g(X)=\frac{1}{4}\int\limits_{-\infty}^{\infty} dV\, \text{sech}^2\,\left(\frac{t V}{2}\right) \,
\left[\tilde{q}_t(X,V)\right]^2 .
\end{equation}
For given $g(X)$ (a monotonic function) and $t$, Eq.~(\ref{LD090A}) is the Schr\"{o}dinger equation for the wave function $\tilde{q}_t(X,V)$ of a quantum particle with mass $m=1/2$ and energy $-V$ moving in the potential $U(X)$. The factor $t^2\gg 1$ in front of the square brackets plays the role of $1/\hbar^2$. Employing the small parameter $1/t$, we will solve Eq.~(\ref{LD090A}) in the WKB 
approximation. As we will see, under some condition that we will specify, the WKB approximation holds for all $X\in (-\infty, \infty)$ except in a small vicinity of the (unique) ``classical turning point" of the ``particle" $X=a(V)$. The turning point is defined by the equality $U(a)+V=0$. Let us introduce the classical momentum of the ``particle",
\begin{equation}\label{classmomentum}
p(X,V)=\sqrt{-V-U(X)} = \sqrt{-V-X-2g(X)}.
\end{equation}
It is a (positive) real function of $X$ in the classically allowed region $X<a$ and a purely imaginary function in the classically
forbidden region $X>a$. The wave function oscillates
in the classically allowed region, and decays exponentially in the classically forbidden region.
The general form of the WKB solution is well known \cite{LL,Bender}:
\begin{numcases}
{\tilde{q}_t(X,V)\simeq} \frac{C_t(V)}{\sqrt{p(X,V)}}\, \cos \left[t\int_{a(V)}^X  p(X',V) dX' -\frac{\pi}{4}\right], & $X<a$,\label{allowed} \\
\frac{C_t(V)}{2\sqrt{|p(X,V)|}}\,\exp\left[-t \int_{a(V)}^X |p(X',V)| dX'\right], & $X>a$. \label{forbidden}
\end{numcases}
To determine the function $C_t(V)$, we use the boundary condition (\ref{E020a}). This yields
\begin{equation}\label{Climit}
 C_t(V)  = \frac{1}{\sqrt{\pi}} \,
  \lim_{X\to +\infty} \exp\left[t\left(\int_a^X|p (X',V)| \,dX'-\frac{2}{3}(X+V)^{3/2}\right)\right] ,
\end{equation}
which can be rewritten as
\begin{equation}\label{Cint}
 C_t(V) = \frac{1}{\sqrt{\pi}} \,
\exp\left[t\left(\int_a^{-V}|p| \,dX'+\int_{-V}^{+\infty}\left(|p|-\sqrt{X+V}\right) dX \right)\right].
\end{equation}
The second integral in the right hand side of Eq.~(\ref{Cint}) converges at $+\infty$ because $\Psi_t(s)$ rapidly goes to zero as $s\to +\infty$ [see Eq.~(\ref{Psilarger})] and therefore $g(X)$ rapidly goes to zero as $X\to \infty$. Now we should plug the asymptotic solutions~(\ref{allowed}) and (\ref{forbidden}) into Eq.~(\ref{LD080a}) and solve the resulting equation for $g(X)$. Continuing to use the large parameter $t\gg 1$, we make the following simplifications:
\begin{itemize}
\item We neglect in Eq.~(\ref{Cint}) an exponentially small contribution of $g(X)$ to the integral in the region of $X>0$ and obtain
\begin{equation}\label{simpleC}
C_t(V)|_{V>0}\simeq \frac{1}{\sqrt{\pi}}\exp\left[t \left(\int_{a(V)}^{0}|p(X,V)|\, dX-\frac{2}{3}V^{3/2}\right)\right] .
\end{equation}
\item We neglect small contributions to the integral in Eq.~(\ref{LD080a}) which come from (i) the classically forbidden region
$X>a$ and (ii) the small non-WKB region around the classical turning point $X=a$.
\item For $r\sim t^{2/3} \gg 1$, the dominant contribution to the integral~(\ref{LD080})  comes from the region of
$t^{1/3} v\gg 1$. Correspondingly, the dominant contribution to the integral~(\ref{LD080a}) comes from the region of
$tV \gg 1$. Therefore, we can approximate $\text{sech}^2\,(tV/2) \simeq 4 \,e^{-tV}$ at $V>0$ and neglect an exponentially small contribution from the region $V<0$.
\item We replace the rapidly oscillating  factor $\cos^2(\dots)$ in Eq.~(\ref{LD080a}), coming from Eq.~(\ref{allowed}),  by $1/2$.
\end{itemize}
As a result, Eq.~(\ref{LD080a}) takes the form of  a formidable-looking nonlinear integral equation for $g(X\leq 0)$:
\begin{equation}\label{E070a}
g(X) = \frac{1}{2\pi} \int\limits_0^{-X-2g(X)} \frac{dV}{\sqrt{|V+X+2g(X)|}}\,e^{t \left(2\int_a^0\sqrt{V+X+2g(X)}\,dX-\frac{4}{3} V^{3/2}-V\right)},\quad X\leq 0.
\end{equation}
Its solution, however, is amazingly simple and, as we announced earlier, independent of $t$:
\begin{equation}\label{gresult}
g(X)=\frac{1}{\pi^2}\left(\sqrt{1-\pi^2 X}-1\right), \quad X\leq 0.
\end{equation}
Miraculously, this $g(X)$ not only ``kills" the $t$-dependent exponent in Eq.~(\ref{E070a}),
\begin{equation}\label{kills}
2\int_{a(V)}^0\sqrt{V+X+2g(X)}\,dX-\frac{4}{3} V^{3/2}-V=0,
\end{equation}
but also solves the remaining equation
\begin{equation}\label{external}
g(X) = \frac{1}{2\pi} \int\limits_0^{-X-2g(X)} \frac{dV}{\sqrt{-V-X-2g(X)}},\quad X\leq 0.
\end{equation}
For the WKB approximation to be valid, we must demand that the characteristic WKB action be large \cite{LL,Bender}:
$$
t \left[\int_{a(V)}^0\sqrt{V+X+2g(X)}\,dX\right] \gg 1.
$$
Using Eq.~(\ref{kills}), we can rewrite this condition as
\begin{equation}\label{WKBcrit1}
t\left(\frac{2}{3}V^{3/2}+\frac{1}{2} V\right)\gg 1.
\end{equation}
Further, for the WKB solution to give a dominant contribution to the integral over $V$ in Eq.~(\ref{LD080}),
the strong inequality (\ref{WKBcrit1}) must hold for $V=-X-2 g(X)$, the upper integration bound in
Eq.~(\ref{external}). For $|X|\ll 1$ we obtain $-X-2g(X)\simeq (\pi^2/4) X^2$, and the applicability condition
is $t X^2 \gg 1$,
or $|r|\gg t^{1/6}\gg 1$.
For $|X|\sim 1$ the applicability condition is simply $t\gg 1$.

Going back to Eq.~(\ref{ansatzPsi}), we see
that $\Psi_t(r)$ is a self-similar function of its arguments:
\begin{equation}\label{E080}
\Psi_t(r)=\frac{t^{2/3}}{\pi^2} \left(-1+\sqrt{1-\frac{\pi^2\, r}{t^{2/3}}}\right)\, .
\end{equation}

Now we are in a position to evaluate $Q_t(s)$ from Eq.~(\ref{Qdirect}). As $\Psi_t(r>0)\simeq 0$, we can write
\begin{equation}\label{Qdirect1}
 - \ln  Q_t(s) \simeq \int_s^{0} dr (s-r) \Psi_t(r) = t^2 \Phi_{-}\left(\frac{s}{t^{2/3}}\right),\quad (-s)\gg t^{1/6},
\end{equation}
where
\begin{equation}\label{Phi}
 \Phi_{-}(z) = \int_z^{0} dX (X-z) g(X) = \frac{4}{15\pi^6}\, \left(1-\pi^2\, z\right)^{5/2}-\frac{4}{15\pi^6}+\frac{2}{3\pi^4}\, z-\frac{1}{2\pi^2}\, z^2 .
\end{equation}
This leads to the exact LDF we are after:
\begin{equation}\label{E110}
-\ln {\cal P}_t(H)\big|_{-H\gg \sqrt{t}\gg 1}=t^2 \Phi_{-}\left(\frac{H}{t}\right) .
\end{equation}

\section{Tale of two tails}

The leading-order $-z\ll 1$ asymptote $\Phi_-(z) \simeq - z^3/12$ yields the height distribution
\begin{equation}\label{TW}
-\ln {\mathcal P}_t(H)\simeq \frac{|H|^3}{12t}, \quad |H|\ll t.
\end{equation}
Although the WKB approximation demands $|H|\gg t^{1/2}$, the leading-order result (\ref{TW}) actually holds under
a weaker condition
$|H|\gg t^{1/3}$, because it coincides  with the left tail of the Tracy-Widom distribution that describes typical fluctuations of height at long times.  The asymptote (\ref{TW}) was obtained in Ref. \cite{DMS}. Furthermore, the authors of Ref. \cite{DMS} arrived at a conclusion
that this asymptote holds at $|H|\sim t$. This conclusion is  in contradiction with our exact large-deviation function (\ref{Phi}) and (\ref{E110}) \cite{wrongscaling}.

The leading-order $-z\gg 1$ asymptote of $\Phi_{-}(z)$ is $\Phi_-(z) \simeq 4|z|^{5/2}/(15 \pi)$. Correspondingly, the $|H|\gg t$ asymptote of the height distribution is the following:
\begin{equation}\label{5over2}
-\ln {\mathcal P}_t(H)\simeq \frac{4 |H|^{5/2}}{15\pi \,t^{1/2}},\quad |H|\gg t.
\end{equation}
This asymptote was obtained in Ref.
\cite{KMS} by using the OFM, and in Ref. \cite{DMRS} in the short-time limit $t<<1$.
As it is clear now, the tail (\ref{5over2}) is present at all times $t>0$. This tail is independent
of the diffusion coefficient $\nu$ \cite{KMS}. Indeed, in the physical variables one obtains
\begin{equation}\label{DF}
-\ln {\mathcal P}_t(H)\simeq \frac{4 \sqrt{2|\lambda|}}{15\pi D}\,\frac{|H|^{5/2}}{t^{1/2}},\quad\quad |H|\gg \frac{ |\lambda|^3 D^2 t}{\nu^4}.
\end{equation}
Therefore, we will call this far-tail asymptote `diffusion-free'. For comparison, the tail~(\ref{TW}) in the physical variables is
\begin{equation}\label{NDF}
-\ln {\mathcal P}_t(H)\simeq \frac{2\,\nu^2|H|^3}{3\,|\lambda|D^2\, t}, \quad\quad
\left(\frac{|\lambda|D^2 t}{\nu^2}\right)^{1/3}\!\!\ll |H|\ll  \frac{|\lambda|^3 D^2 t}{\nu^4}.
\end{equation}
Here too the KPZ nonlinearity dominates over the diffusion, but the tail still depends on $\nu$.

The exact LDF~(\ref{E110}) describes a smooth crossover between the Tracy-Widom tail (\ref{TW})
and the far tail (\ref{5over2}) in the region of $|H|\sim t$. For reference purposes, we present more accurate small- and large-$|z|$ asymptotics:
\begin{equation}
{\Phi_{-}(z) =}
 \label{more}
\begin{cases}
-\frac{1}{12}\,z^3-\frac{\pi ^2}{96}\, z^4 -\frac{\pi ^4}{320}\, z^5 - \dots,   & -z\ll 1, \\
 \frac{4}{15 \pi}\,|z|^{5/2} -\frac{1}{2 \pi^2}\,z^2-\frac{2}{3 \pi^3}\,|z|^{3/2}+\dots,  & -z\gg 1 .
\end{cases}
\end{equation}

\section{Numerical evaluation}
\label{NUM}

The probability distribution of $H$ can be extracted from the exact generating function (\ref{LD050}) and (\ref{LD410}) \cite{CDR}. It is equal to
\begin{equation}\label{P[g]}
{\mathcal P}_t(H)=\int_{-\infty}^{\infty}d u\, e^{H-t^{1/3}u}\exp(-e^{H-t^{1/3}u})\,G_{t}(u)\,,
\end{equation}
where $G_{t}$ is given by the difference of two Fredholm determinants,
\begin{equation}\label{gt}
G_{t}(u)=\det[I-\hat{P}_u (\hat{B}_t - \hat{A})\hat{P}_u]-\det[I-\hat{P}_u \hat{B}_t\hat{P}_u]\,.
\end{equation}
The operators $\hat{A}$ and $\hat{B}_t$ have respective kernels $A(x,x')=\mbox{Ai}(x)\mbox{Ai}(x')$ and
\begin{equation}\label{Bt}
B_{t}(x,x')=\int_{0}^{\infty} d v\left[\frac{\mbox{Ai}(x+v)\mbox{Ai}(x'+v)}
{1-e^{-t^{1/3}v}}+\frac{\mbox{Ai}(x-v)\mbox{Ai}(x'-v)}{1-e^{t^{1/3}v}}\right]\,.
\end{equation}

\begin{figure}
  \includegraphics[width=100mm]{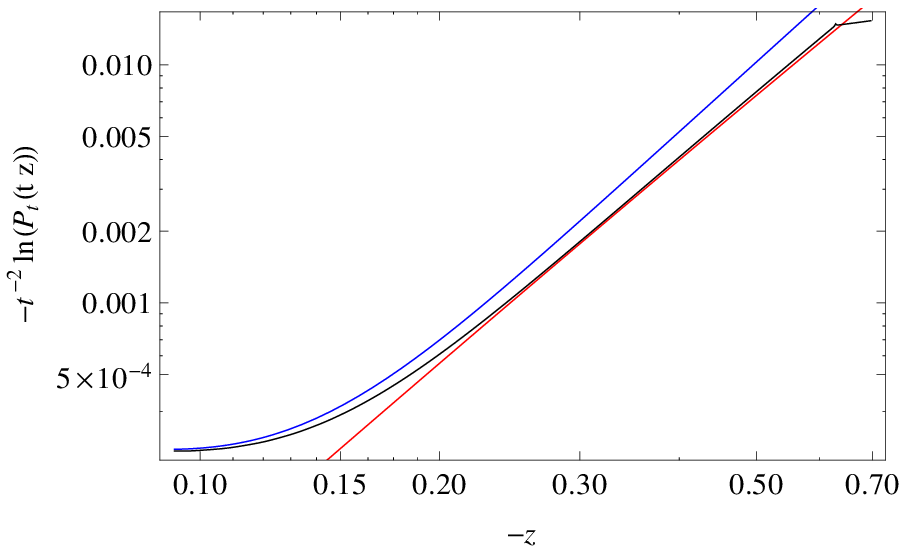}
  \includegraphics[width=100mm]{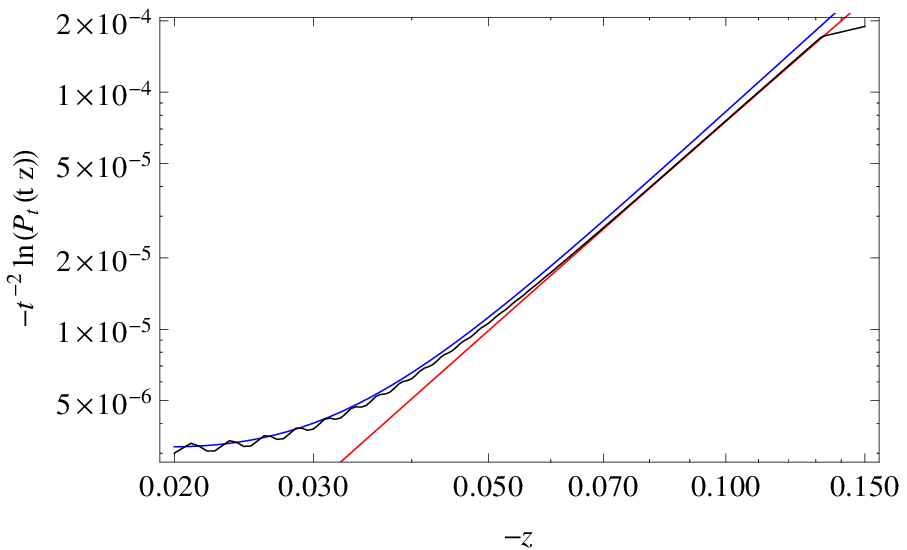}
  \caption{Log-log plot of $-t^{-2}\ln{\mathcal P}_t(t z)$ versus $-z$ for $t=100$ (top) and $t=1000$ (bottom). The black curve corresponds to the numerical evaluation of Eq.~(\ref{P[g]}) described in Sec. \ref{NUM}. The spurious oscillations at small $|z|$ in the bottom plot result from a crude discretization of the integral over $u$ in Eq.~(\ref{P[g]}). The red curve is the exact large deviation function $\Phi_{-}(z)$ from Eq.~(\ref{Phi}). The blue curve is the Tracy-Widom asymptotics $-t^{-2}\ln(t^{-1/3}f(t^{2/3}z))$ computed from Eq.~(\ref{TW left tail}), which takes into account the pre-exponential factor.}
  \label{fig P t=1000}
\end{figure}

The central part of ${\mathcal P}_t(H)$, corresponding to typical fluctuations, was computed numerically in Ref. \cite{PS2011} using the method introduced by Bornemann in \cite{B2010} for accurate evaluations of Fredholm determinants. Here we push the computations further in order to reach the left tail of ${\mathcal P}_t(H)$.

Bornemann's method consists in approximating a Fredholm determinant $\det[I-\hat{L}]$ by evaluating the multiple integrals in the Fredholm expansion by Gauss-Legendre quadrature with $M$ points, which is exact for integrands of degree at most $2M-1$, and converges exponentially fast with $M$ quite generally. The approximate Fredholm expansion with discretized integrals can then be resummed as a single determinant, and one has
\begin{equation}\label{det approx}
\det[I-\hat{L}]\simeq \det[\delta_{\ell,\ell'}+\sqrt{w_{\ell}w_{\ell'}}\,L(x_{\ell},x_{\ell'})]_{\ell,\ell'=1,\ldots,M}\;.
\end{equation}
For Gauss-Legendre quadrature the points $x_{\ell}$ are the zeroes of the $M$-th Legendre polynomial
$$
P_{M}(x)=(2^{M}M!)^{-1}\partial_{x}^{M}(x^{2}-1)^{M}\,,
$$
and the corresponding weights $w_{\ell}$ are given by
$$
w_{\ell}=\frac{2}{MP_{M-1}(x_{\ell})P_{M}'(x_{\ell})}\,.
$$
An additional step is needed if the kernel $L$ has infinite support, since Gauss-Legendre quadrature requires integrals on a finite segment. This can be remedied by a change of variables $A(\varphi(y),\varphi(y'))$ in the kernel.

\begin{figure}
  \includegraphics[width=100mm]{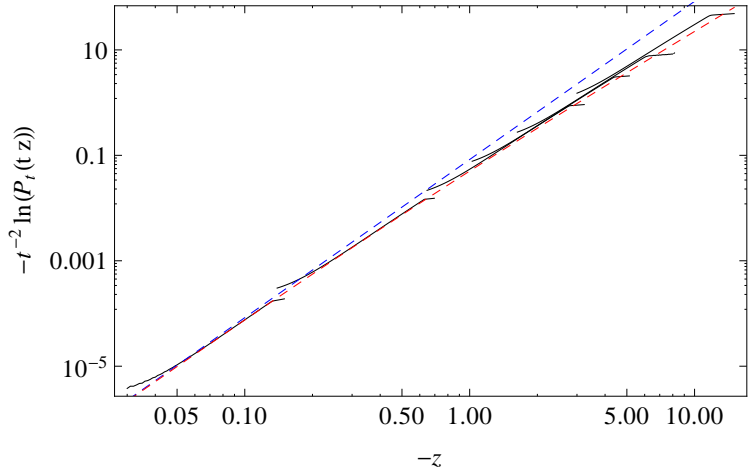}
  \caption{Log-log plot of $-t^{-2}\ln{\mathcal P}_t(t z)$ versus $-z$ for $t=1,2.5,5,10,100$ and $1000$. The black curves corresponds to the numerical evaluation of Eq.~(\ref{P[g]}) described in Sec.~\ref{NUM}, with longer times toward the left. The horizontal plateaux at the right end of each curve are artefacts due the the finite value of the number of points of discretization $M$. The dashed red curve is the exact large deviation function $\Phi_{-}(z)$ from Eq.~(\ref{Phi}). The dashed blue curve is the Tracy-Widom asymptotics $|z|^{3}/12$.}
  \label{fig P t=1-1000}
\end{figure}

An additional difficulty in the application of Bornemann's method to Eq.~(\ref{P[g]}) is that the kernel $B_{t}$ is itself given by an integral (\ref{Bt}). We also evaluate this integral by Gauss-Legendre quadrature, after a change of variables $v=\varphi(y)$ which maps the interval $[0,\infty)$ to a finite segment. We used $\varphi(u)=10\tan(\pi u/2)$ for the Gauss-Legendre quadrature of both the Fredholm determinants and the kernel $B_{t}$.

The computation of the left tail of ${\mathcal P}_t(H)$ is much more demanding than the computation of the central part of the distribution \cite{PS2011}, where it was sufficient to use $M=30$ and double-precision numbers. In order to go deeper into the left tail, we had to evaluate $G_{t}(u)$ for larger negative values of $u$, for which the approximation (\ref{det approx}) of the Fredholm determinants in (\ref{gt}) converges more slowly as $M$ increases. Besides, the oscillations of $G_{t}(u)$ for $u<0$ lead to cancelations in the integration over $u$ in Eq.~(\ref{P[g]}), and require higher floating-point precision. Both issues of course increase the computation time. We found that $M=150$ and floating-point numbers with $150$ digits was a good compromise between how far to the tail we could go and how long the computation would take. With these parameters, each value of $G_{t}(u)$ took about $8$ hours with `Mathematica' \cite{Wolfram} on a single core of a personal computer. The integral over $u$ in (\ref{P[g]}) is then evaluated by simple rectangular quadrature between $u=-15$ and $u=10$ with step $\delta u=0.25$.

With the numerical scheme described above, we evaluated the left tail of ${\mathcal P}_t$ for $t=1, 2.5, 5, 10, 100$ and $1000$. The results are plotted in Figs.~\ref{fig P t=1000} and~\ref{fig P t=1-1000} alongside with the exact LDF~$\Phi_{-}(z)$ from Eq.~(\ref{Phi})
and the  Tracy-Widom asymptotic. The agreement between the numerical results and the exact LDF is rather good.  As one can see from  Fig.~\ref{fig P t=1000}, a deviation from the Tracy-Widom asymptotic appears already at quite small $|z|$, and this deviation is well described by the exact $\Phi_{-}(z)$.

\section{Discussion}

We derived exact LDF of height of the 1+1 KPZ equation with the droplet initial condition at long times for $\lambda H<0$. This LDF, see Eqs.~(\ref{Phi}) and (\ref{E110}), describes a smooth crossover
from the Tracy-Widom distribution tail  at small $|H|/t$, which scales as $|H|^3/t$,
to a diffusion-free tail at large $|H|/t$, which scales as $|H|^{5/2}/t^{1/2}$. The diffusion-free tail exists at all times $t>0$, but it is ``pushed" to larger and larger $|H|$ as time grows.

Le Doussal \text{et al.} \cite{DMS} argued that, at long times,  models in the KPZ universality class exhibit a third-order
phase transition from a strong-coupling to a weak-coupling phase. Their argument was based on Eq.~(\ref{TW}).  Here we have shown that the asymptotic~(\ref{TW}) is not valid at $-H\sim t$. Still, their interpretation of the large deviations of height in terms of a third-order phase transition holds. Indeed, sufficiently close to the ``critical point" $H=0$ one still has
\begin{equation}
{\lim_{t\to \infty} -\frac{1}{t^2}\,\ln {\mathcal P}(H=zt,t) =}
 \label{thirdorder}
\begin{cases}
z^3/12 , & 0<-z\ll 1, \\
 0,  & z>0.
\end{cases}
\end{equation}
In the light of our results, at $|H|/t\gg 1$, the strong-coupling phase becomes diffusion-free. Here the height fluctuations are dominated by a \emph{large-scale} optimal noise history \cite{KMS}.

The diffusion-free tails $\sim |H|^{5/2}/t^{1/2}$ at very large negative $\lambda H$ have been also  obtained with the OFM for the KPZ equation in $1+1$ dimensions with other types of initial conditions \cite{KK2007,KK2009,MKV,KMS,JKM}, including the flat and stationary initial conditions.  It would be interesting to reproduce them from exact representations of the height distribution at long times.

Finally, the KPZ universality class is defined in terms of \emph{typical} fluctuations at long times. It should not come as a surprise, therefore, that statistics of \emph{large deviations} are in general different among different models belonging to the KPZ universality class.

\section*{ACKNOWLEDGMENTS}

B.M. acknowledges financial support from the Israel Science Foundation (grant No. 807/16).


\bigskip\bigskip

\begin{thebibliography}{99}

\bibitem{KPZ}  M. Kardar, G. Parisi, and Y.-C. Zhang, Phys. Rev. Lett. \textbf{56}, 889 (1986).

\bibitem{HHZ} T. Halpin-Healy and Y.-C. Zhang, Phys. Reports \textbf{254}, 215 (1995); T. Halpin-Healy and K. A. Takeuchi,
J. Stat. Phys. \textbf{160}, 794 (2015).

\bibitem{Krug}
J. Krug, Adv. Phys. \textbf{46}, 139 (1997).

\bibitem{Corwin}
I. Corwin, Random Matrices: Theory Appl. \textbf{1}, 1130001 (2012).

\bibitem{QS}
J. Quastel and  H. Spohn, J. Stat. Phys. \textbf{160}, 965 (2015).

\bibitem{S2016}
H. Spohn, arXiv:1601.00499.

\bibitem{experiment1} W. M. Tong and R.W. Williams, Annu. Rev. Phys. Chem. \textbf{45}, 401 (1994);
L. Miettinen, M. Myllys, J. Merikoski, and J. Timonen, Eur. Phys. J. B \textbf{46}, 55 (2005),
M. Degawa, T. J. Stasevich, W. G. Cullen, A. Pimpinelli, T. L. Einstein, and E. D. Williams,
Phys. Rev. Lett. \textbf{97}, 080601 (2006).


\bibitem{CLD} P. Calabrese, and P. Le Doussal,  Phys. Rev. Lett. \textbf{106}, 250603 (2011);
P. Le Doussal and P. Calabrese, J. Stat. Mech. P06001 (2012).

\bibitem{SS} T. Sasamoto, H. Spohn, Phys. Rev. Lett. \textbf{104}, 230602 (2010).

\bibitem{CDR} P. Calabrese, P. Le Doussal, A. Rosso, Europhys. Lett.
\textbf{90}, 20002 (2010).

\bibitem{Dotsenko} V. Dotsenko, Europhys. Lett. \textbf{90}, 20003 (2010).

\bibitem{ACQ} G. Amir, I. Corwin, and J. Quastel, Comm. Pur. Appl. Math.
\textbf{64}, 466 (2011).

\bibitem{IS} T. Imamura and T. Sasamoto, Phys. Rev. Lett. \textbf{108}, 190603 (2012); J.
Stat. Phys. \textbf{150}, 908 (2013).

\bibitem{Borodinetal} A. Borodin, I. Corwin, P.L. Ferrari, and B. Vet\H{o}, Mathematical Physics, Analysis and Geometry \textbf{18}, 1 (2015).

\bibitem{TWGOE} C. A. Tracy and H. Widom, Comm. Math. Phys. \textbf{177}, 727 (1996).

\bibitem{TW} C. A. Tracy and H. Widom, Comm. Math. Phys. \textbf{159}, 151 (1994).

\bibitem{DIK2008} P. Deift, A. Its and I. Krasovsky, Commun. Math. Phys. \textbf{278}, 643 (2008).

\bibitem{BR} J. Baik and E.M. Rains, J. Stat. Phys. \textbf{100}, 523 (2000).

\bibitem{experiment2} K.A. Takeuchi and M. Sano, Phys. Rev.
Lett. \textbf{104}, 230601 (2010); J. Stat. Phys. \textbf{147}, 853–890 (2012),
K. Takeuchi, M. Sano, T. Sasamoto, and H. Spohn,  Sci. Rep. \textbf{1}, 34 (2011); K. A. Takeuchi,
Phys. Rev. Lett. \textbf{110}, 210604 (2013); T. Halpin-Healy and Y. Lin, Phys. Rev. E
\textbf{89}, 010103R (2014).

\bibitem{DMS} P. Le Doussal, S. N. Majumdar, and G. Schehr,  
 EPL \textbf{113}, 60004 (2016).

\bibitem{DMRS} P. Le Doussal, S. N. Majumdar, A. Rosso, and G. Schehr, 
Phys. Rev. Lett. \textbf{117}, 070403 (2016).

\bibitem{Fogedby1998} H.C. Fogedby, Phys. Rev. E \textbf{57}, 4943 (1998).

\bibitem{Fogedby1999} H.C. Fogedby, Phys. Rev. E \textbf{59}, 5065 (1999).

\bibitem{Fogedby2009} H.C. Fogedby and W. Ren, Phys. Rev. E \textbf{80}, 041116 (2009).

\bibitem{KK2007} I. V. Kolokolov and S. E. Korshunov, Phys. Rev. B \textbf{75}, 140201(R) (2007).

\bibitem{KK2008} I. V. Kolokolov and S. E. Korshunov, Phys. Rev. B \textbf{78}, 024206 (2008).

\bibitem{KK2009} I. V. Kolokolov and S. E. Korshunov, Phys. Rev. B \textbf{80}, 031107 (2009).

\bibitem{MKV} B. Meerson, E. Katzav and A. Vilenkin, Phys. Rev. Lett. \textbf{116}, 070601 (2016).

\bibitem{KMS} A. Kamenev, B. Meerson and P.V. Sasorov, Phys. Rev. E  \textbf{94}, 032108 (2016).

\bibitem{JKM} M. Janas, A. Kamenev and B. Meerson,  Phys. Rev. E \textbf{94}, 032133 (2016).

\bibitem{signlambda}
Note that Refs. \cite{MKV,KMS,JKM} assumed $\lambda<0$. Changing the sign of $\lambda$ is equivalent to changing the sign of $h$.

\bibitem{Hairer} M. Hairer, Annals of Math. \textbf{178}, 559 (2013).

\bibitem{different} Note that Ref. \cite{KMS} used a different rescaling of the variables. As a result, their LDF of height -- the action $S(H)$ -- is related to $\Phi(H)$ via $S(H)=8 \,\Phi(-H/2)$.

\bibitem{DLMF} NIST Digital Library of Mathematical Functions, http://dlmf.nist.gov/9.7.E5 .


\bibitem{LL} L.D. Landau and E.M. Lifshitz, \textit{Quantum Mechanics} (Pergamon Press, London, 1965), Chapter VII.

\bibitem{Bender} C.M. Bender and S.A. Orszag, \textit{Advanced Mathematical Methods for Scientists and Engineers. Asymptotic Methods and Perturbation Theory} (McGrow-Hill, New York, 1978), Chapter 10.

\bibitem{wrongscaling} The authors of Ref. \cite{DMS} made an \emph{a priori} assumption that the function $q_t(r,v)$ exhibits the following scaling behavior at $t\to \infty$:
$q_t(r,v)\simeq t^{1/3} \phi (r/t^{2/3}, vt^{1/3})$. It follows from our results that this assumption is incorrect already at $-s\gtrsim t^{1/6}$.

\bibitem{PS2011} S. Prolhac and H. Spohn, Phys. Rev. E \textbf{84}, 011119 (2011).

\bibitem{B2010} F. Bornemann, Math. Comp., \textbf{79}, 871 (2010); Markov Processes Relat. Fields \textbf{16}, 803 (2010).

\bibitem{Wolfram} Wolfram Research, Inc., Mathematica, Version 8.0.4.0, 2011.




\end{thebibliography}
\end{document}